\documentclass[twocolumn]{aa}
\usepackage{graphicx}
\usepackage{txfonts}
\begin{document}
\title{Using globular clusters to test gravity in the weak acceleration regime:
NGC 7099\thanks{Based on observations collected at the 
European Southern Observatory, Chile (ESO No. 71.D-0311 and 075.D-0043).
The radial velocities used in this work are only available in electronic form
at the CDS via anonymous ftp to cdsarc.u-strasbg.fr (130.79.128.5)}}

   \author{R. Scarpa\inst{1} \and G. Marconi\inst{1} \and R. Gilmozzi\inst{2} 
 \and G. Carraro\inst{3}
          }

   \offprints{R. Scarpa; rscarpa@eso.org}

\institute{European Southern Observatory, 
Santiago, Chile \and European Southern Observatory, Garching bei M\"unchen, Germany
\and Universit\`a di Padova, Italy; Universidad de Chile, Santiago}

\date{}
\authorrunning{Scarpa, Marconi, Gilmozzi, and Carraro}
\titlerunning{Testing gravity in the weak acceleration regime}

\abstract{}{A test of Newton's law of gravity in the low acceleration
regime using globular clusters is presented and new results for the core
collapsed globular cluster NGC 7099 given.}{The run of the
gravitational potential as a function of distance is probed by studying
the velocity dispersion profile of the cluster, as derived from a set
of 125 radial velocities with accuracy better than 1 km s$^{-1}$.  The
velocity dispersion profile is traced up to $\sim 18$ pc from the
cluster center.}  {The dispersion is found to be maximal at the
center, then decrease until $10 \pm 2$ pc from the center, well inside
the cluster tidal radius of 42 pc. After that the dispersion remains
basically constant with an average value of $2.2 \pm 0.3$ km $s^{-1}$. Assuming
a total V mag of M(V)=$-7.43$ mags for NGC 7099, the acceleration at $10\pm 2$ pc
from the center is $ 1.1^{+0.4}_{-0.3}\tau \times 10^{-8}$ cm
s$^{-2}$, where $\tau$ is the mass-to-light ratio. Thus, for $\tau \la
2$ typical of globular clusters, the flattening of the velocity
dispersion profile occurs for a value of the internal acceleration of
gravity that is fully consistent with $a_0=1.2\times 10^{-8}$ cm s$^{-2}$ observed in
galaxies. }  {This new result for NGC 7099 brings to 4 the clusters
with velocity dispersion profile probing acceleration below $a_0$. All
four have been found to have a flat dispersion profile at large radii
where the acceleration is below $a_0$, thereby mimicking elliptical galaxies
qualitatively and quantitatively.  Whether this indicates a failure
of Newtonian dynamics in the low acceleration limit or some more
conventional dynamical effect (e.g., tidal heating) is still
unclear. However, the similarities emerging between very different
globular clusters, as well as between globular clusters and elliptical
galaxies, seem to favor the first of these two possibilities.

  }

\maketitle
%

\section{Introduction}

The gravitational accelerations governing the dynamics of cosmic
structures are typically orders of magnitude smaller than the one
probed in our laboratories or in the solar system.  Thus, any time
Newton's law is applied to galaxies (e.g., to infer the existence of
dark matter), its validity is severely extrapolated.  Although there
should be no reason to distrust Newton's law in the weak
acceleration regime, unanimous agreement has been reached (e.g.,
\cite{binney04}) on the fact that galaxies start deviating from
Newtonian dynamics, and dark matter is needed to reconcile
observations with predictions, $always$ for the same value of the
gravitational acceleration $a_0\sim1.2\times 10^{-8}$ cm s$^{-2}$
(\cite{begeman91}), as computed considering only baryons.  This
systematics, more than anything else, suggests that we may be facing a
breakdown of Newton's law rather than the effects of dark matter.

If for a moment we assume that Newtonian dynamics breaks down in the
low acceleration limit and non-baryonic dark matter does not exist,
then the behavior of galaxies in the low acceleration limit should be
typical of many other systems, as long as the acceleration is the
same. According to this idea we looked at globular clusters, the
largest virialized structure believed not to contain dark matter.
Being free falling toward the Milky Way, globular clusters are
only affected by tidal stress, which is in most cases well below
$a_0$. Therefore the internal dynamic of globular clusters do probe
the same range of accelerations probed by galaxies, making them ideal
for testing gravity in the low acceleration regime without the
complication of dark matter.

In previous papers we studied the dynamical properties of $\omega$
Centauri (\cite{scarpa03}), M15, and NGC 6171 (Scarpa, Marconi \&
Gilmozzi 2004A,B), where it was shown that the velocity dispersion
remains constant at large radii as soon as the acceleration reaches
$a_0$, as is the case for elliptical galaxies (e.g., \cite{mehlert00})--
a puzzling and potentially important result.  To further
generalize it, we present here new results for NGC 7099, a compact
cluster located at 8.0 kpc from the sun and 7.1 kpc from the Galactic
center (\cite{harris96}).

\begin{figure}
\centering
\includegraphics[height=8cm]{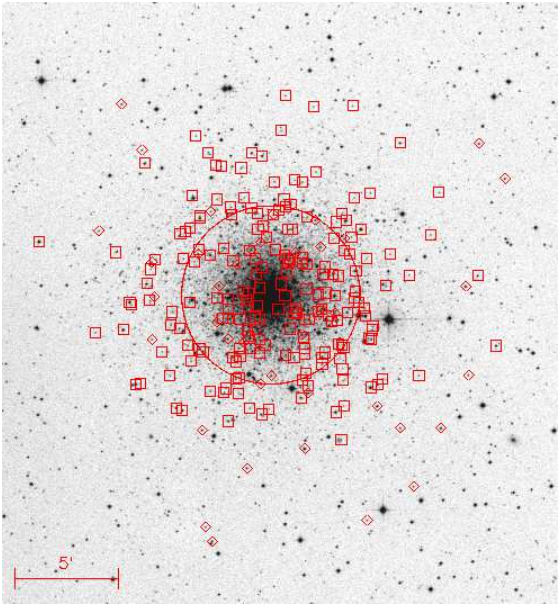}
\caption{\label{figIma} The cluster NGC 7099 as seen in the digital
sky survey with overplotted the location of the 234 stars for which
a radial velocity with error smaller than 5 km s$^{-1}$ was derived. Squares
mark cluster members, while diamonds are non-members.  The cluster
center was set to 21$^{h}$40$^{m}$22$^{s}$
$-23^{\circ}$10$^{'}$45$^{''}$ (\cite{harris96}). North is up and east
to the left. The circle has 10 pc radius.}
\end{figure}

\section{Observation and data analysis}

The initial selection of targets was based on color, as derived from
the analysis of ESO Imaging Survey (EIS) frames. After identifying the
giant branch and main sequence of the cluster in the H-R diagram, we
prepared a catalog of about 300 targets, mostly in the sub-giant
branch and turn off, and apparent V magnitude $16<$V$<19$.
Observations were then obtained with FLAMES (\cite{pasquini02}) at the
ESO VLT telescope. FLAMES is a fiber multi-objects spectrograph,
allowing the simultaneous observation of up to 130 objects. We
selected the HR9B setup that includes the magnesium triplet covering
the wavelength range $5143 < \lambda < 5346$ \AA\ at resolution
R=25900. Stellar astrometry was derived by cross correlating the stellar
positions on the EIS frame with coordinates from the US Naval
Observatory (USNO) catalog, which proved to have the required accuracy
(0.3 arcsec) for FLAMES observations. Three different  fiber
configurations were used in order to cover the
external region of the cluster as much as possible (Fig. 1). For each configuration,
three 2700 s exposures were obtained under good atmospheric conditions
(clear sky and seeing $\sim 1$ arcsec). The first set of images was
obtained during period 71 on December 5, 2003, and the other two
during period 75 on August 29 and 30, 2005.

Data reduction was performed within IRAF, using standard reduction
procedures. After extraction and wavelength calibration, all spectra
were cross correlated with respect to the target with the best
spectrum.  The  three configurations shared a
small number of stars, to evaluate and eliminate possible
offsets in the velocity zero point. 
A posteriori, we verified that no correction was necessary
down to a level of accuracy of 300 m s$^{-1}$, well below the
accuracy required for our study.  Finally, keeping in mind that we
are only interested in the velocity dispersion, 
the global velocity zero point was simply
derived by identifying a few lines in the spectrum of the template.  In
total, 234 radial velocities (all velocities presented here are
heliocentric) with accuracy better than 5 km s$^{-1}$ were obtained.
 
Final membership was assigned according to the observed radial
velocity. In total 194 members were identified without ambiguity in
the velocity space (Fig. 2) at a radial velocity of $\sim -185$ km
s$^{-1}$, that is fully consistent with the quoted velocity for this cluster
of $-185.9\pm 0.6$ km s$^{-1}$ (\cite{gebhardt95}). All 
cluster members have error on the radial velocity smaller than 2.5 km
s$^{-1}$, with 125 members having error smaller than 1 km
s$^{-1}$. The larger errors, up to 5 km s$^{-1}$, refer to field stars
that have a significantly different spectral energy distribution from
the one of the template, degrading the accuracy of the cross
correlation.

Finally, in NGC 7099 we found no evidence of
ordered rotation to the level of 0.75 km s$^{-1}$.

\begin{figure}
\centering
\includegraphics[height=7.5cm]{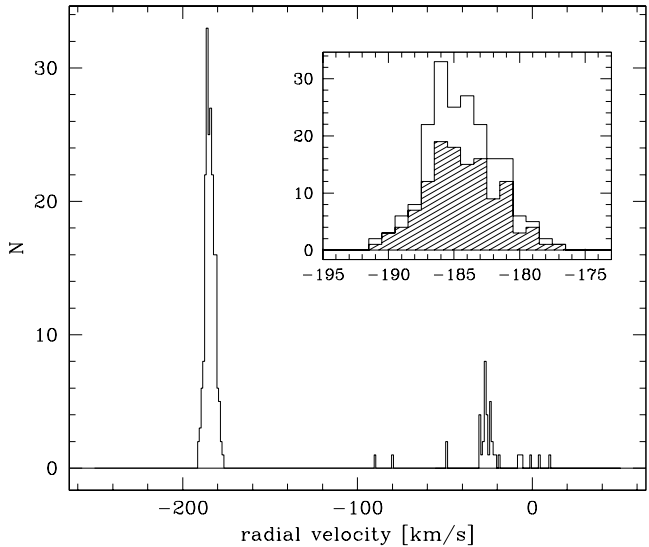}
\caption{\label{figHis} Distribution of the 234 radial velocities with
error smaller than 5 km s$^{-1}$. The position of the cluster members is
evident at v$\sim -185$ km s$^{-1}$.  The inset shows  the
distribution of the 194 cluster members. The shaded area corresponds to
the 125 stars with errors on the radial velocity smaller than 1 km s$^{-1}$.
These were used in figures 3 and 4 to derive the velocity dispersion.}
\end{figure}

\begin{figure}
\centering
\includegraphics[height=6cm]{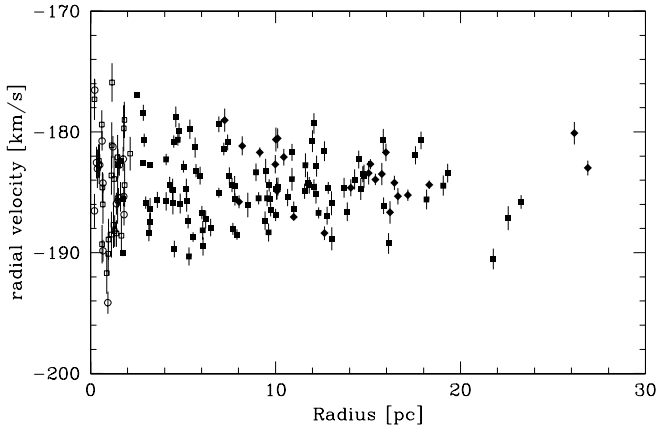}
\caption{\label{figVel} The distribution of the 125 radial velocities
with error smaller than 1 km s$^{-1}$. Objects observed during period 71
(solid diamonds) and period 75 (solid squares) are shown with
different symbols.  Data from  Zaggia et al. (1992; open circle) and
Gebhardt et al. (1995; open squares) covering the central region of the
cluster are also shown.}
\end{figure}

\section{Velocity dispersion profile of NGC 7099}

Our cluster member velocity measurements have uncertainties
between 0.5 and 2.5 km s$^{-1}$. Thus, to avoid artificially
inflating the dispersion we are trying to measure, we computed the
dispersion profile for subsets of our data, each one with increasing
maximum error. Statistically indistinguishable dispersion profiles
were found, with the higher accuracy on the velocity balanced by the
smaller size of the sample.  Here we present the dispersion profile
as derived from the 125 members with accuracy on the radial velocity
better than 1 km s$^{-1}$. These stars are distributed over $2 < r <
27$ pc from the cluster center.  To cover the very central region of
the cluster we included two data sets from the literature.  The first
refers to 16 stars within 50 arcsec from the center (\cite{zaggia92}).
These velocities, which have average error of about 0.9 km s$^{-1}$,
have been shifted by 2.5 km s$^{-1}$ to match our average radial
velocity.  The second data set is from a sample of 132 velocities
(\cite{gebhardt95}), from which we selected the 27 velocities with
errors smaller than 2 km s$^{-1}$.  The data cover the central 1
arcmin and have been shifted by 1.5 km s$^{-1}$ to match our data. The
larger error associated with these data is not a problem because, at
the center, the dispersion is significantly larger than this. As a
whole the data from the literature smoothly match our data in the
region of overlap (Fig. 3), allowing us to build a well-sampled
velocity dispersion profile from the center to almost 20 pc
(Table 1). Beyond that we found only 5 cluster members, providing
little or no information at all on the velocity dispersion at these
distances (Fig. 4).

Looking at figures 3 and 4, we see no indication of the expected
Keplerian fall off of the velocity dispersion at large radii, but rather,
the dispersion seems to remain noticeably constant if not increasing
toward the very end of the probed region.  Excluding the very last
point in Fig. 4, that due to the huge associated uncertainty is
consistent with any conceivable model, we conclude that the dispersion
converges beyond r$ = 10 \pm 2$ pc toward the asymptotic values of
$2.2 \pm 0.3$ km s$^{-1}$. Note that the flattening occurs well inside
the cluster tidal radius of 42 pc (\cite{harris96}).

\begin{figure*}
\centering
\includegraphics[height=7cm]{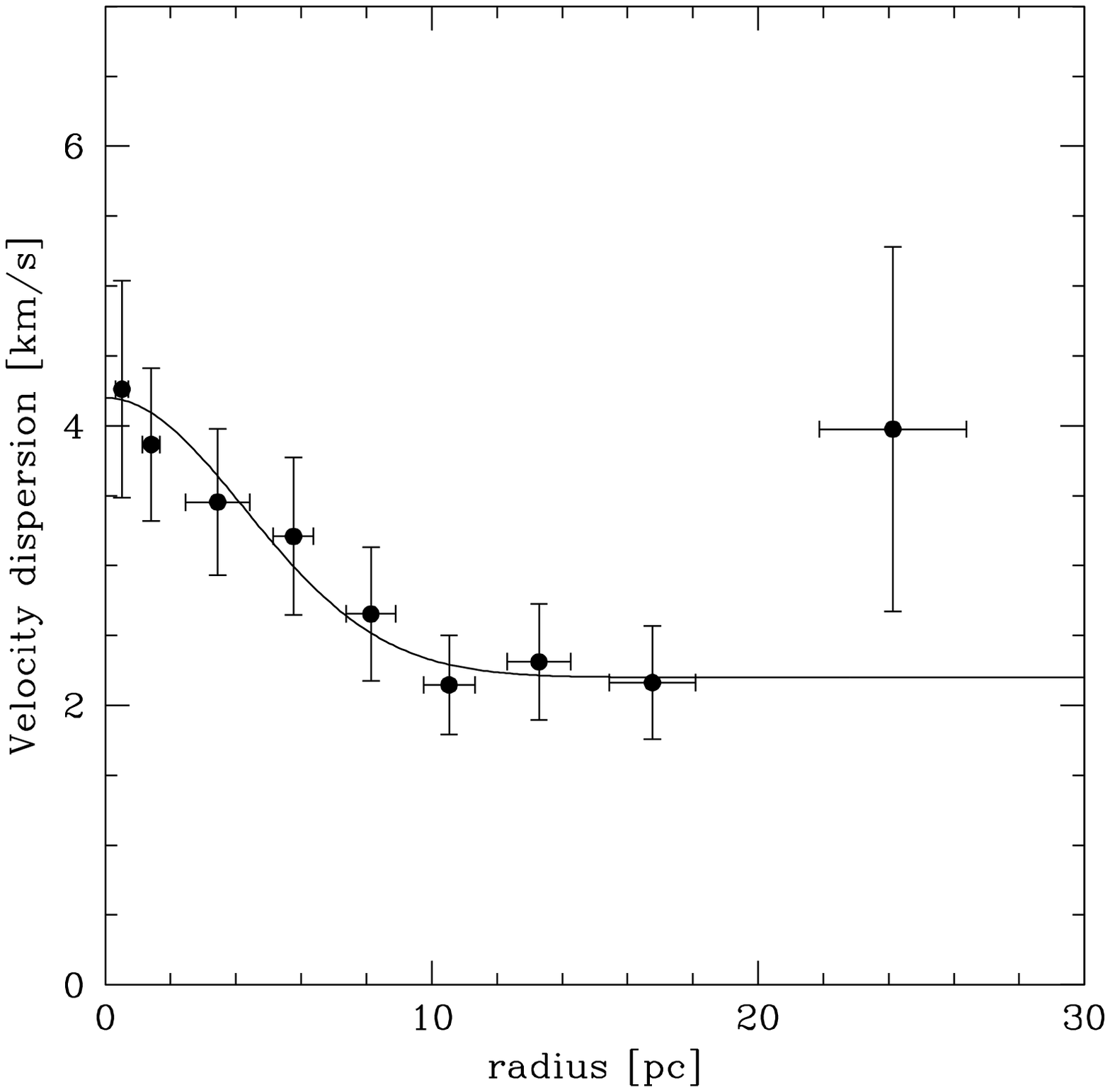}\includegraphics[height=7cm]{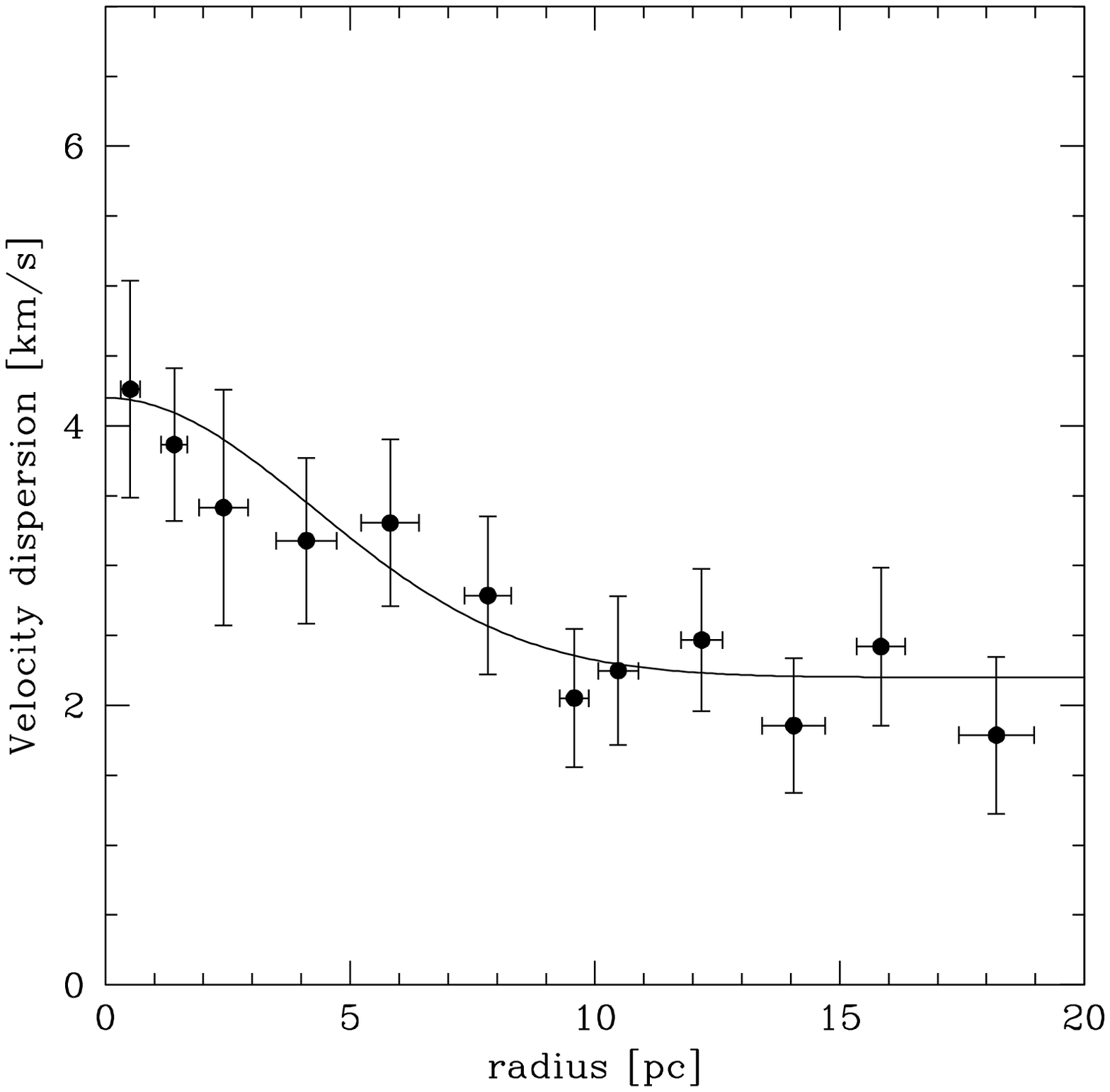}
\caption{\label{figDis} {\bf Left.} The radial velocity dispersion profile as
derived from the 125 radial velocities with error smaller than 1 km s$^{-1}$,
plus 16 velocities from Zaggia et al. (1992) and 27 from
Gebhardt et al. (1995). Data from the literature contributes only to the
two innermost points.  The abscissa of each point is the average of
the points in the bin.  Error bars in the x direction show the
$1\sigma$ dispersion of the data in the bin, while error bars in the y
direction represent the $1\sigma$ uncertainty on the dispersion.  
The solid line (a Gaussian plus a
constant) is not a fit to the data but is meant only as a guide for the
eye. {\bf Right.} Enlargement of the previous figure showing the profile 
with different  binning.}
\end{figure*}

\begin{table}[b]
\caption{Radial velocity dispersion of NGC 7099}
\begin{tabular}{cccc}
Bin limits (pc)  & Stars/bin & bin center (pc) & $\sigma$ (km s$^{-1}$)\\
\hline 
$  0-0.9    $&$    16  $&$      0.51 $&$  4.26 \pm     0.78 $\\
$  0.9-1.8  $&$    27  $&$      1.45 $&$  3.87 \pm     0.55 $\\
$  1.8-4.8  $&$    24  $&$      3.44 $&$  3.45 \pm     0.52 $\\
$  4.8-7.0  $&$    18  $&$      5.76 $&$  3.21 \pm     0.56 $\\
$  7.0-9.5  $&$    18  $&$      8.14 $&$  2.65 \pm     0.48 $\\
$  9.5-12   $&$    23  $&$     10.53 $&$  2.15 \pm     0.35 $\\
$  12-15    $&$    19  $&$     13.29 $&$  2.31 \pm     0.41 $\\
$  15-20    $&$    18  $&$     16.76 $&$  2.16 \pm     0.40 $\\
$  >20      $&$     5  $&$     24.13 $&$  3.97 \pm     1.30 $\\
\hline
\hline
\end{tabular}
\end{table}

\section{Discussion}

Within errors the velocity dispersion profile of NGC 7099 is flat for
r$ > 10 \pm 2$ pc. Assuming a total absolute magnitude of M$_V =
-7.43$ (\cite{harris96}), the mass of NGC 7099 is $7.8\tau \times
10^4$ M$_\odot$, where $\tau$ is the mass-to-light ratio.  A simple
look at Fig. 1 shows that virtually all the mass of the cluster is
contained well within 10 pc. Thus, whatever the mass distribution in
NGC 7099 might be, the internal acceleration of gravity at this radius
is a = GM$/r^2 = 1.1^{+0.4}_{-0.3}\tau \times 10^{-8}$ cm s$^{-2}$.
Theoretical considerations and direct estimates of the mass to
light ratio in globular clusters indicate $\tau \la 2$ (e.g.,
\cite{mclaughlin05}). The flattening therefore occurs at an
acceleration that is in reasonable agreement with what is observed in
galaxies. It is indeed remarkable that given a bit of information from
galaxies rotation curves, $a_0$, one can predict within a factor 2 the
flattening of the dispersion profile in a globular cluster, a
structure 3 order of magnitude smaller and 6 order of magnitude less
massive than a typical galaxy.

As a whole, the new data for NGC 7099 brings to four the number of
globular clusters for which the dispersion profile is seen to flatten
out below $a_0$.  Specifically, in all four cases the velocity
dispersion is maximal at the center, then rapidly declines to converge
toward a constant value at large radii where $a<a_0$, instead of
decreasing according to a Keplerian falloff.  These four clusters have
different physical properties (mass and size), as well as different
dynamical and evolutionary histories, nevertheless they behave exactly
in the same way, also quantitatively and qualitatively mimicking the
behavior of high surface brightness elliptical galaxies {e.g.,
\cite{mehlert00}).

It might be argued that these 4
clusters are all at comparable galactocentric distances and
therefore are at present experiencing similar tidal stresses and external
fields. This might be why they behave similarly.  However, the
dynamical history is different, and comparable tidal stresses are
deemed to produce very different effects on clusters of different
size and mass, thus a non negligible amount of fine tuning is necessary
to explain our observations.  

The alternative, which is what we are trying to test, is that the
flattening might be due to a breakdown of Newtonian dynamics. It is
well known that the flattening of the dispersion profile
and/or the rotation curve occurs at $a_0$ in galaxies.  This is at the base of a
particular modification of Newtonian Dynamics known as MOND
(\cite{milgrom83}), capable of successfully fitting the properties of
a large number of stellar systems without invoking the existence of
non-baryonic dark matter (\cite{mcgaugh98}; \cite{sanders02};
\cite{milgrom03}; \cite{scarpa06}).  According to MOND, however, the
velocity dispersion in NGC 7099 should remain constant as
soon as the total, i.e. internal plus external, acceleration of gravity is
$a\la a_0$. At the distance of NGC 7099, the acceleration of gravity
due to the Milky Way is $v^2/r\sim 2\times10^{-8}$ cm s$^{-2}$,
assuming a rotational velocity of 220 km $s^{-1}$ for the Galaxy.
Thus, the total acceleration is somewhat above $a_0$, and according to
MOND as originally stated (\cite{milgrom83}), only minor deviation
from the Newtonian prediction should occur.  Nevertheless -- and whatever
MOND might predict -- in NGC 7099 and the other clusters we have studied so
far, something happens to the velocity dispersion at $a_0$, and this
cannot be due to dark matter as it allegedly is in galaxies.  Whether
this indicates something as fundamental as a breakdown of Newtonian
dynamics is unclear.  In this respect, it is worth noting that the
four globular clusters studied so far probe the typical accelerations
observed in high surface brightness galaxies (and they behave like
them). By contrast, the velocity dispersion profile of low surface
brightness elliptical galaxies is remarkably flat (e.g.,
\cite{mateo97}; \cite{wilkinson06}).  These galaxies, believed to be
dark matter dominated all the way to their center, do have internal
acceleration everywhere below $a_0$.  Thus, if the parallel with
galaxies holds, then also low-concentration globular clusters with
acceleration everywhere below $a_0$ should have constant velocity
dispersion.  This firm prediction will be the subject of the next
paper of this series; but in the meanwhile, we conclude by saying that both
alternatives remain viable, although the evidence of a breakdown
of Newtonian dynamics in the low acceleration limit is growing.

\acknowledgements
We thank R. Falomo, T. Richtler, and R. Capuzzo Dolcetta 
for useful discussions and suggestions, 
and the referee D. Mc Laughin for his useful report that helped in
improving the manuscript.

\end{document}